\def\apj{ApJ}
\def\aps{ApJS}
\def\anr{ARA\&A}
\def\aaa{A\&A}

\def\nat{Nat}

\def\mn{MNRAS}
\def\et{et al.}

\def\msu{M$_{\odot}$}
\def\kms{km s$^{-1}$ }

\input psfig
\documentstyle[aaspp]{article}

\begin{document}

\title{Abell 3627: A Nearby, X-ray Bright, and Massive Galaxy Cluster}

\author{H. B\"ohringer, D.M. Neumann,
S. Schindler\altaffilmark{1}}
\affil{Max-Planck-Institut f\"ur extraterrestrische Physik, D-85740 Garching,
Germany}
\author{R.C. Kraan-Korteweg}
\affil{Observatoire de Paris, DAEC, Unit\'e associ\'ee au CNRS, D0173,
et \`a l'Universit\'e Paris 7, 92195 Meudon, France}

\altaffiltext{1}{also at Max-Planck-Institut f\"ur Astrophysik,
D-85740 Garching,Germany}

\begin{abstract}
The cluster A3627 was recently recognized to be a very massive,
nearby cluster
in a galaxy survey close to the galactic plane. We are reporting
on ROSAT PSPC observations of this object which confirm that the cluster is indeed
very massive. The X-ray emission detected from the cluster extends over
almost 1 degree in radius. The X-ray image is not
spherically symmetric and shows indications of an ongoing cluster merger.
Due to the strong interstellar absorption the spectral analysis and the
gas temperature determination are difficult. The data are consistent with an
overall gas temperature in the range 5 to 10 keV. There are signs of
temperature variations in the merger region.

A mass estimate based on the X-ray data yields values of
$0.4 - 2.2 \cdot 10^{15}$ \msu \ if extrapolated to
the virial radius of $3 h_{50}^{-1}$ Mpc.
In the ROSAT energy band (0.1 - 2.4 keV) the cluster emission yields a flux
of about $2 \cdot 10^{-10}$ erg s$^{-1}$ cm$^{-2}$ which makes A3627
the 6$^{th}$ brightest cluster in the ROSAT All Sky Survey.
The cluster was missed in earlier X-ray surveys because it was confused
with a neighbouring X-ray bright, galactic X-ray binary (1H1556-605).
The large X-ray flux makes A3627 an important target for future studies.
\end{abstract}


\section{Introduction}

In a deep galaxy survey  behind the Milky Way ($|b| \le 10\deg $)
the cluster A3627 was recently found by Kraan-Korteweg et al. (1996) to be
a very rich, nearby cluster of galaxies rivaling the prominent
Perseus and Coma clusters in its mass and galaxy content. The
observed recession velocity of 4882 km, s$^{-1}$ corresponding to
a velocity of 4655  km s$^{-1}$ with respect to the Local Group places
A3627 at a distance of about 93 Mpc
(we are using a Hubble constant of $H_0 = 50$ km s$^{-1}$
Mpc$^{-1}$ in this paper). Therefore the cluster is
one of the most prominent mass concentrations
in the local Universe. Kraan-Korteweg \et\ (1996)
found a galaxy velocity dispersion
of 903 km s$^{-1}$ and attributed a mass of $5 \cdot 10^{15}$~\msu\ to
this cluster.
Although A3627 has been cataloged by Abell, Corwin \&
Olowin (1989) as a richness class 1 cluster with 59 galaxies
it attracted little attention in the past {\sl because} of its proximity
to the galactic plane ($l = 325\deg $, $b = -7\deg $). It has also not been
detected in previous X-ray surveys (e.g. Piccinotti \et\ 1982,
Kowalski \et\ 1984, Lahav \et\ 1989).

We have observed this prominent cluster with the ROSAT PSPC. The results of this
observation support the finding of Kraan-Korteweg \et\ (1996) that A3627 is a
very massive cluster. In this paper we give a detailed report on the X-ray
properties of A3627 and describe its structure as revealed by the ROSAT
images. Section 2 provides an account of the X-ray morphology. The X-ray
spectral analysis is discussed in Section 3. Section 4 describes the mass
determination. In Section 5 these results are discussed and
Section 6 provides a summary.

\section{X-ray Morphology}

The cluster A3627 was observed with the ROSAT PSPC in September 1992 and
March 1993 for a total exposure time of 11,257 sec. The galactic hydrogen
column density at this location is quite high (around
$1.8\cdot 10^{21}$ cm$^{-2}$). Therefore no photons are detected in the
soft energy band of ROSAT below the carbon edge (0.1 to 0.4 keV).
An X-ray image produced from the hard band counts (0.5 - 2 keV)
is shown in Fig. 1.
The X-ray emission of the cluster fills almost the whole field of
view of the detector. The total ROSAT PSPC count rate in the (0.5 to 2.0
keV band) is found to be $ 7.0 \pm 0.3 $ cts s$^{-1}$. The X-ray emission 
was integrated out to a radius of 51.5 arcmin and the background was taken
from the remaining area outside. In addition we also analysed regions
inside this radius to the north-east and south-west of the cluster and
found that these areas give background values consistent with those from
the outer anulus. This comparison shows that uncertainties in the 
vignetting correction at the very outer edge of the PSPC field do not
affect the results. Anyway these uncertaitnies are expected to
be more critical in the soft energy channels and less in the hard part
of the ROSAT energy window used here.
For a distance of
93 Mpc this corresponds to  a flux of $2.1 (\pm 0.3) \cdot 10^{-10}$ erg
s$^{-1}$ cm$^{-2}$ and to an X-ray luminosity of
$2.2 (\pm 0.3) \cdot 10^{44}$ erg s$^{-1}$
in the ROSAT energy band (0.1 - 2.4 keV).
For these calculations a galactic hydrogen column density of
$1.4 - 2.0 \cdot 10^{21}$ cm$^{-2}$ and gas temperatures in the range of
5 - 9.5 keV are assumed and the conversion factors are determined using
a Raymond-Smith code (Raymond \& Smith 1977) with a metallicity of
0.35 in solar units.
Furthermore, inspecting
the cluster image in the ROSAT All Sky Survey, we find no
significant additional X-ray flux beyond a radius of 1 degree.
The total cluster count rate in the 0.5 - 2.0 energy band of
$6.8 \pm 0.3$ cts s$^{-1}$ determined from
the All Sky Survey data is consistent with the result of the pointed
observation.

As can be seen in Fig. 1, A3627 is clearly not spherically symmetric.
The cluster shows a strong elongation in the direction of PA $\sim
130\deg $ (measured from north over east).
The central maximum of the X-ray emission is at the position
16h14 22, -60d52 20 (J2000) and a secondary X-ray maximum at
16h14 10, -60d50 34. The bright
point source about 23 arcmin to the north-west of the cluster
center, at the position 16h11 51.7, -60d37 44.0 (J2000), originates from a
Seyfert galaxy which was found in the course of the galaxy survey
(Woudt \et\ 1996). The galaxy was classified as Seyfert 1 and with a velocity
of 4711 km~s$^{-1}$ it certainly is a cluster member. Another point source,
best visible in Fig. 3, is coincident with the radio galaxy
PKS1610-60 at the position 16h15 03.7, -60 54 26 (J2000), one of the
three central galaxies of the cluster.
This source is used
to check the attitude of the ROSAT pointing, which is found to be correct
within an uncertainty of 3 arcsec.
We detect a series of
other point sources in the PSPC field of view, that will be described in
a follow-up publication.

To see if a varying absorbing hydrogen column density is partly
responsible for the observed morphological features,
we have studied the hardness variation of the X-ray emission across the cluster
image. For this study we mainly
use the hardness ratio of the energy bands R4 (PSPC channel 52 - 69) and
R5 (PSPC channel 60 - 90). The ROSAT PSPC energy bands are defined by
Snowden \et\ (1994). These energy bands are closest to the
absorption cut-off for the relevant range of column density values and therefore
this hardness ratio is sensitive to a variation of the absorbing
column density. We do not detect any significant variation of the
hardness ratio across the cluster image, however, except that the hardness
ratio is much lower for the background than for the cluster.

Assuming that there is an approximately spherically symmetric main component
dominating the cluster image, we analyse the surface brightness
profile of the western part of the cluster. We use the data from a sector
centered on the main maximum (16h14 22, -60d52 20) within the angular region
from  -180\deg to 45\deg \ (extending from south to north-east
in counter-clockwise direction).
This region is opposite to
the prominent elongation in the cluster and appears therefore less
distorted.
The surface brightness profile is shown in Fig.~2. We fit a
$\beta$-model (e.g. Cavaliere \& Fusco-Femiano 1976;
Jones \& Forman 1984) to the surface
brightness profile, $S(r)$, (see Fig. 2), using the formula:
\begin{equation}
S(r) = S_0 \left( 1 +{r^2 \over r_c^2} \right) ^{-3\beta +1/2}  .
\end{equation}

Since the core radius of the cluster is so much larger than
the resolution of the ROSAT PSPC with a half power radius of
20 arcsec for the point spread function, a convolution
with the point spread function does not affect the fitting
results significantly and was therefore neglected. The X-ray
background is taken as a free fit parameter in this analysis and even though
there is not much area at the true background level in the field of view,
the fit yields a background value consistent with the one determined from the
outermost ring of the PSPC.  
The resulting fit parameters
and their uncertainties are: central surface brightness, $S_0 =
10^{-2}$ counts s$^{-1}$ arcmin$^{-2}$,  $\beta = 0.555~
(+0.045,-0.035)$, and core radius, $r_c = 9.95 \pm 1.0$ arcmin
($262 \pm 24~h_{50}^{-1}$ kpc).
The core radius for the X-ray surface brightness is very similar to
the optical core radius of 10.4 arcmin (Kraan-Korteweg \et\ 1996).
The surface brightness
profile keeps increasing inwards 
inside the core radius and therefore the data
points deviate slightly from the $\beta $ model fit
inside a radius of about 5 arcmin. In the outer parts the surface
brightness profile is not very smooth. Both deviations from the $\beta $
model, that generally fits the X-ray surface brightness profile of clusters
quite well, may reflect the disturbances by the ongoing merger in the
cluster as discussed below.

We use the $\beta $ model fitted to the less disturbed side of the
cluster to construct a synthetic spherically symmetric cluster model
that could approximately describe the main body of the cluster. By
subtracting this model cluster image from the observed cluster image
of Fig. 1, a residual image is obtained (Fig. 3).
One can clearly see a very compact cluster component superposed on the
main cluster in the south east.
This is most likely a smaller subcluster in the
process of merging with the main cluster body.
The excess X-ray emission is clearly extended and the X-ray spectrum
is consistent with a thermal spectrum of hot intracluster plasma
(see section 3 and Fig. 6).
This characteristics essentially excludes the interpretation of the
X-ray emission shown in Fig. 3 as a galactic X-ray source like
stars, accretion sources or a supernova remnant.

The morphology displayed in Fig. 3 implies that
the merger has obviously already
progressed to a stage where the smaller subunit is being distorted.
Besides the subcluster there
is some excess emission opposite to the subcluster which may
reflect details in the dynamics of the merging process.

\begin{figure}[h]
\psfig{figure=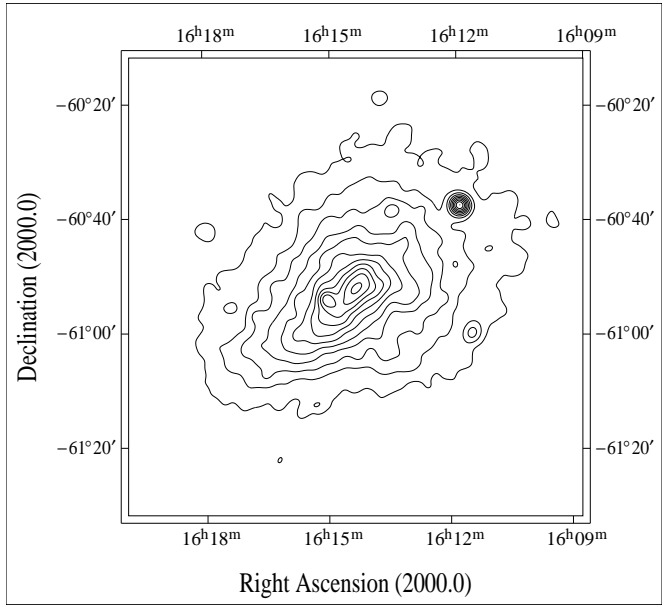,height=8cm}
\caption{ROSAT PSPC image of A3627. Only photons in the hard energy band
(0.5 to 2.0 keV) are used and the image is smoothed by a Gaussian
with  $\sigma $ = 1 arcmin. This count rate image is produced from the 
photon count map by dividing by the exposure map and by correcting
for vignetting of the X-ray telescope. The X-ray background was
not subtracted.  The contour levels start at
$1.6 \cdot 10^{-3}$ cts s$^{-1}$ arcmin$^{-2}$ and increase in steps of
$9.6\cdot 10^{-4}$ cts s$^{-1}$ arcmin$^{-2}$.   }
\end{figure}

\begin{figure}[h]
\psfig{figure=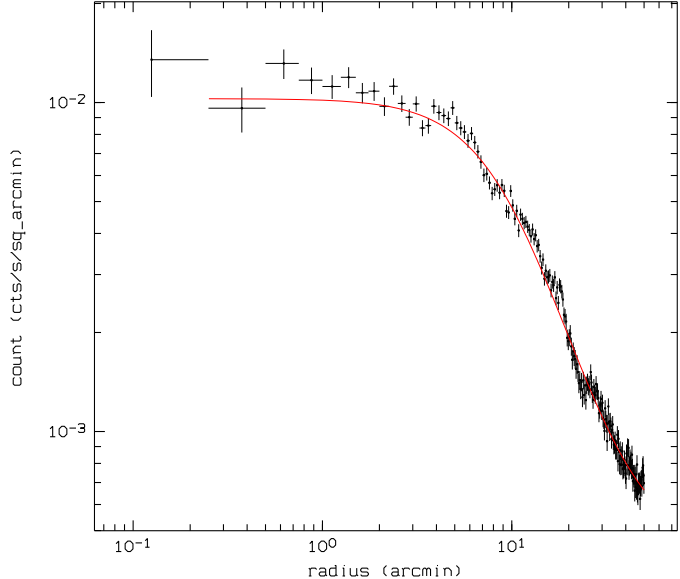,height=8cm}
\caption{X-ray surface brightness profile of A3627 as observed with the ROSAT
PSPC. Only photons received in the hard energy band (0.5 - 2 keV) 
and only data from the region from PA = -180$\deg $ to PA = 45$\deg $ were
	d. The solid line shows the best fitting $\beta $ model
with the parameters $S_0 = 10^{-2}$ cts s$^{-1}$
arcmin$^{-2}$, $\beta = 0.55$ , and $r_c = 9.95$ arcmin.}
\end{figure}

\begin{figure}[h]
\psfig{figure=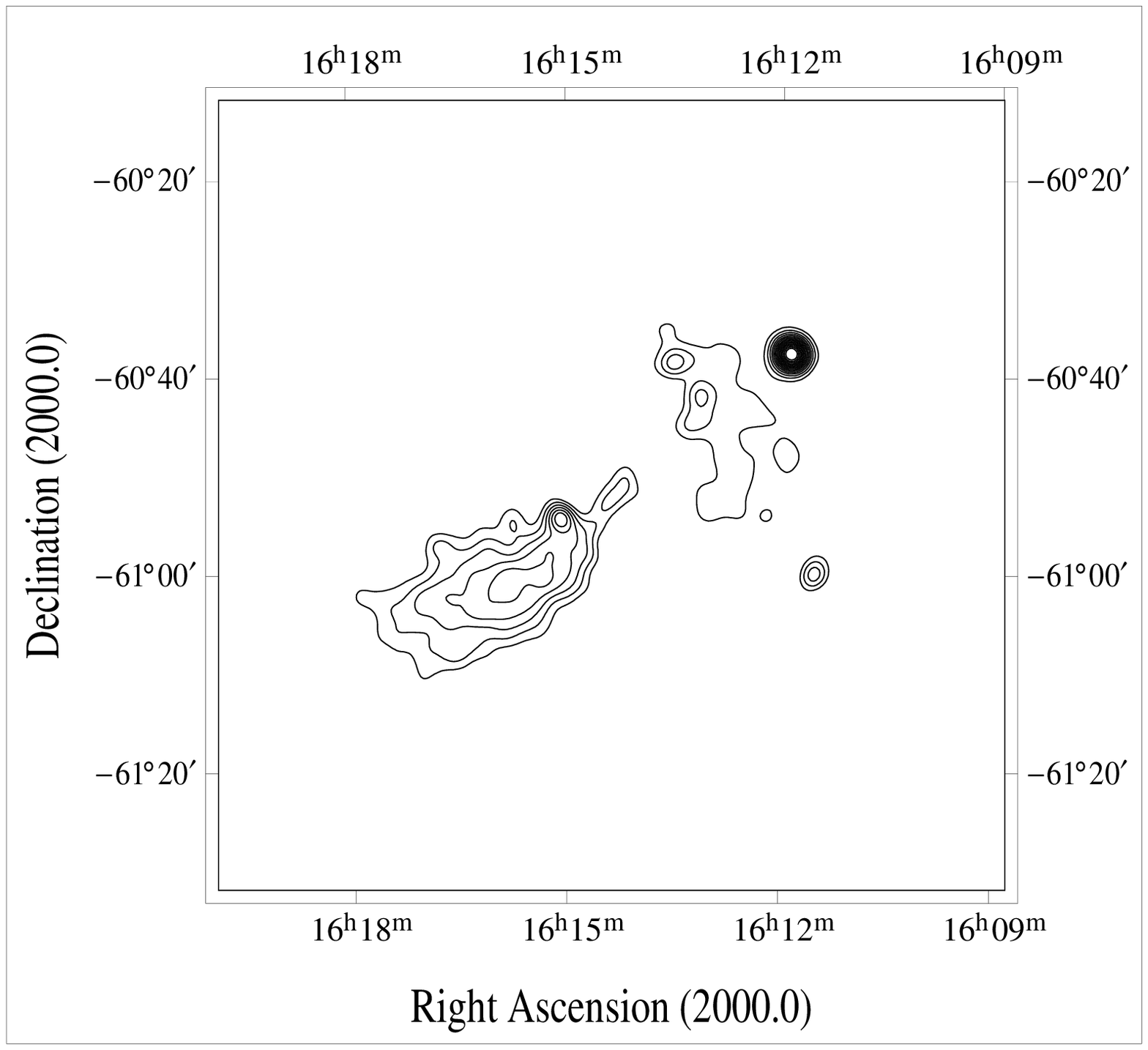,height=8cm}
\caption{Residual X-ray image of A3627 obtained by subtracting
a spherically symmetric model
from Fig. 1.
The parameters for the symmetric model image are taken from a $\beta $
model fit to the less disturbed side of the cluster as described in the text.
The contour levels start at $1.12 \cdot 10^{-3}$ cts s$^{-1}$ arcmin$^{-2}$
and increase in steps of $4.8\cdot 10^{-4}$ cts s$^{-1}$ arcmin$^{-2}$, so that
every second contour corresponds to a contour level in Fig. 1. }
\end{figure}

\eject
\section{X-ray Spectra}

In total about 60,000 photons were received from the cluster X-ray source.
This allows us to derive spectra of good statistical quality
for different regions in the
cluster. Due to the strong absorption in the galactic plane, however,
the leverage to constrain the spectral parameters is very
small. Therefore it is quite difficult to obtain good
constraints for all the spectral parameters simultaneously:
temperature, $T$, hydrogen column density, $N_H$,
and metal abundance. The best strategy to get
reliable parameter constraints is to fit the temperature
and the absorbing column density simultaneously
and to repeat this fitting procedure for the
relevant range of expected metallicities. We use
metallicity values of 0.2, 0.35 and 0.5 of solar abundances (as quoted by
Anders \& Grevesse 1989). We find that the metallicity has little influence on the results
for $T$ and $N_H$. For the spectral analysis we remove the four
brightest point sources in the cluster.

Fig. 4 shows for example the PSPC spectrum for a large region of the
less disturbed part of the cluster. The spectrum is taken from an area
in the sector -180\deg to 90\deg (excluding the south-eastern quadrant)
inside a radius of 10 arcmin.
The strong absorption of the soft part of the spectrum
is evident. The $\chi ^2$ contours for the fit shown in Fig. 5
give the constraints of the temperature and column density and imply
an overall cluster temperature
in the undisturbed part in the range of 5 to 9.5 keV
(for $1\sigma $ errors, with best fitting values around 6 - 7 keV).
Splitting the area in different
subregions does not provide further significant information,
since the errors of
the measurement are always larger than the differences. However, an
indication of a higher temperature in the center prevails. The fits
for the central 5 arcmin region
show no constraint at
the high temperature side and
best fitting values around 10 keV.

Performing a similar analysis in the disturbed part of the cluster,
that is in the sector 90\deg to 180\deg
(the south-eastern quadrant) we find high
temperatures within 5 arcmin of the central region which are of the
same order as in the central region of the undisturbed part.
In the outer parts of the subclump the
temperature is considerably lower (2.5 ${+2\brack -0.5}$ keV) than in the
rest of the cluster.
Fig. 6 shows the $\chi ^2$ contours
for the temperature and column density fit in the subclump for a region
with 10 arcmin diameter, centered
13 arcmin south-east of the center. The difference in the temperature
of this area and of
the overall spectrum has a $\sim 1.5 \sigma$  significance.
Thus there surely are strong and interesting temperature variations
that could provide important
information about the merger physics. They should be studied
with an X-ray observatory with a wider energy window and a higher spectral
resolution, for example ASCA or SAX.

\begin{figure}[h]
\psfig{figure=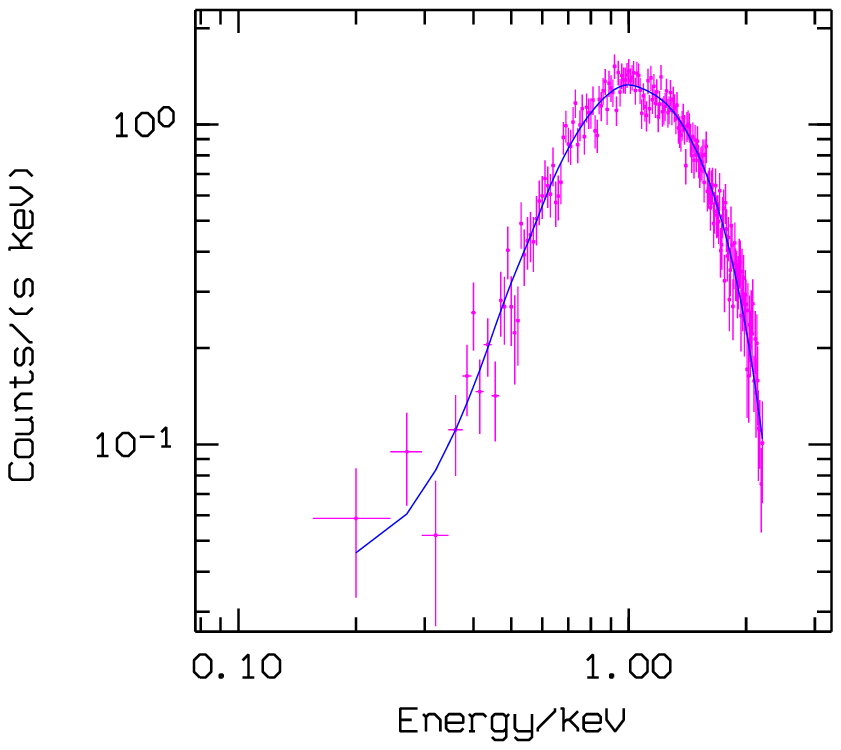,height=8cm}
\caption{X-ray spectrum and fit of a Raymond-Smith model for
the less disturbed part of the cluster A3627. The fact that
hardly any source photons are received below 0.4 keV is a result
of the high absorbing hydrogen column density}
\end{figure}

\begin{figure}[h]
\psfig{figure=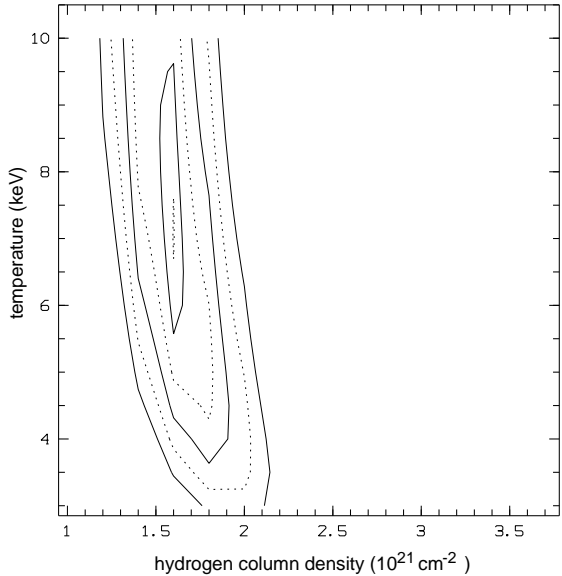,height=8cm}
\caption{{$\chi ^2$} contours for the fit of a Raymond-Smith model to
the X-ray spectrum from the less disturbed part of the cluster out to
a radius of 16.7 arcmin. The temperature and the hydrogen column density
are varied in the fit. The contour levels encircle the {$\chi ^2$}
minimum and indicate the 1, 2, 3, 4, and 5 sigma levels.}
\end{figure}

\begin{figure}[h]
\psfig{figure=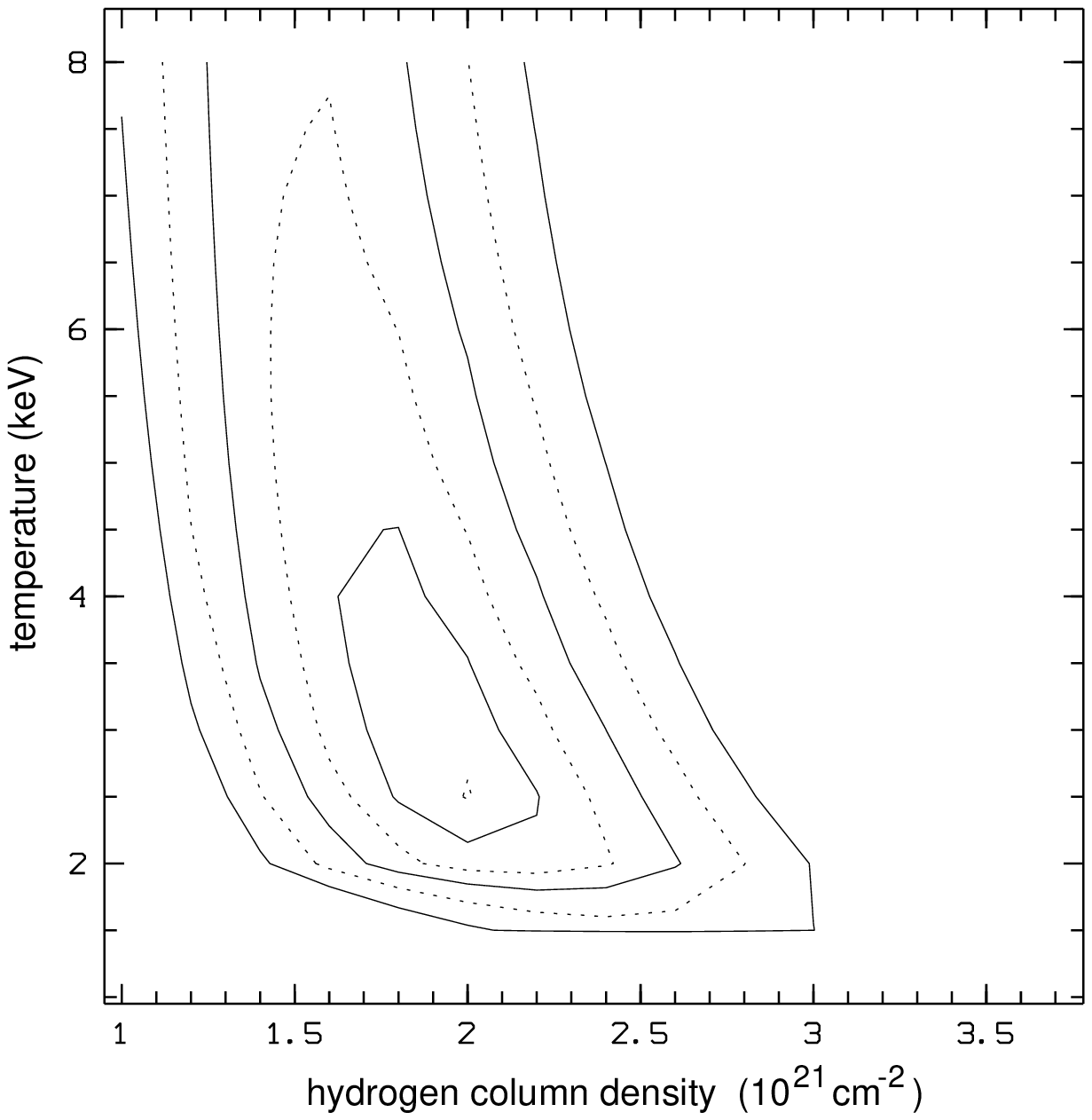,height=8cm}
\caption{{$\chi ^2$} contours for the fit of a Raymond-Smith model to
the X-ray spectrum from a region in the infalling subclump
13 arcmin from the center of A3627.
The temperature and the hydrogen column density
are varied in the fit.
The contour levels encircle the {$\chi ^2$}
minimum and indicate the 1, 2, 3, 4, and 5 sigma levels.}
\end{figure}

\section{Mass Determination}

The above results allow to determine the gas mass distribution
of the main part of the cluster.
We use the fitting results to approximate the gas density
profile. For a $\beta $ model surface brightness profile of the
form given in eq.(1),
the density of the X-ray emitting plasma is given by
\begin{equation}
\rho_g(r) = \rho_{g0} \left( 1 +{r^2 \over r_c^2} \right)
^{-{3\over 2}\beta }  ,
\end{equation}
where $\beta $ is the slope
parameter determining the scale height of the gas density distribution.
The emissivity of the intracluster plasma is almost
independent of the temperature
for the energy band covered by the ROSAT PSPC.
In the temperature range from
2 to 10 keV the emissivity changes only by about 6\%. Therefore
the imperfect knowledge of the gas temperature introduces only
a very small uncertainty in the gas density distribution.

From the gas density distribution and the information on
the intracluster temperature we can obtain a gravitational mass
estimate of the cluster. We  assume that hydrostatic equilibrium
of the intracluster gas
approximately holds in the main part of the cluster. In
fact, Schindler (1996) studied the mass analysis in
N-body/hydrodynamic simulations and showed
that the assumption of hydrostatic
equilibrium leads to very robust results in the mass determination
if obvious regions of disturbances are avoided. The gravitational
mass of the cluster is determined by
\begin{equation}
M(<r) = - {r~kT\over G~\mu m_p} \left( {d \log \rho_g\over d\log r} +
{d \log T \over d\log r} \right)         .
\end{equation}
For the temperature profile we use a range of models: with a given
temperature at the core radius between
9.5 and 5 keV, temperature profiles corresponding to
polytropic models with values of the polytropic index in the range
0.9 to 1.3 are used. The results for the upper and lower limits to the
mass profile are shown in Fig. 7 together with the result for an
isothermal model with $T = 7$ keV and the gas mass profile.

The mass profiles in Fig. 7 are extrapolated to a radius of
$3 h_{50}^{-1}$ Mpc. This should be close to the virial radius of the cluster
as explained below. The X-ray emission is only observed
to a radius of $1.4 h_{50}^{-1}$ Mpc, however.
The values for the mass limits
for these two radii and for a radius of 1 Mpc are listed in Table 1.
The gas mass fractions noted in Table 1 are typical for rich clusters
where one finds values in the range
10 - 30\% for large radii and sometimes smaller values in the
inner regions (e.g. David \et\ 1994; B\"ohringer 1994).
The gas mass distribution can be determined quite precisely with
the ROSAT data with little influence from the uncertainty in the temperature.
Since the values for the gas mass fractions are consistent with other cluster
studies where the spectral data are less affected by absorption,
we conclude that the
mass estimates are basically correct within the given error limits.

The virial radius of a cluster, $r_v $,
with given mass can be calculated analytically
in the limit of the spherical collapse approximation from the formula
(see e.g. Gunn \& Gott 1972, White \et\ 1993):
\begin{equation}
r_v = \left( {M \over 1.38 \cdot 10^{15} h_{50}^{-1} M_{\odot} }
\right) ^{1/2} \Omega^{-0.2} \cdot 3 h_{50}^{-1}~~  Mpc  .
\end{equation}
For a closed Universe we find a virial radius of
$1.7 - 3.8$ Mpc and for $\Omega = 0.4$ values of $2 - 4.5$ Mpc.
For the most likely isothermal model with $T_g = 7$ keV we obtain
$r_v = 2.8 - 3.3$ Mpc for $\Omega = 0.4$ to $1.0$, and a total
mass $M_v \sim 1.2 \cdot 10^{15}$ \msu\ (here the most popular range for the 
value of $\Omega $ is used with a lower bound from the analysis of the large
scale velocity fields (e.g. Dekel 1994) and a closed Universe 
as the upper bound). 
Therefore one should expect that the
virialized part of the cluster system extends to a radius of the
order of 3 Mpc.

\begin{figure}[h]
\psfig{figure=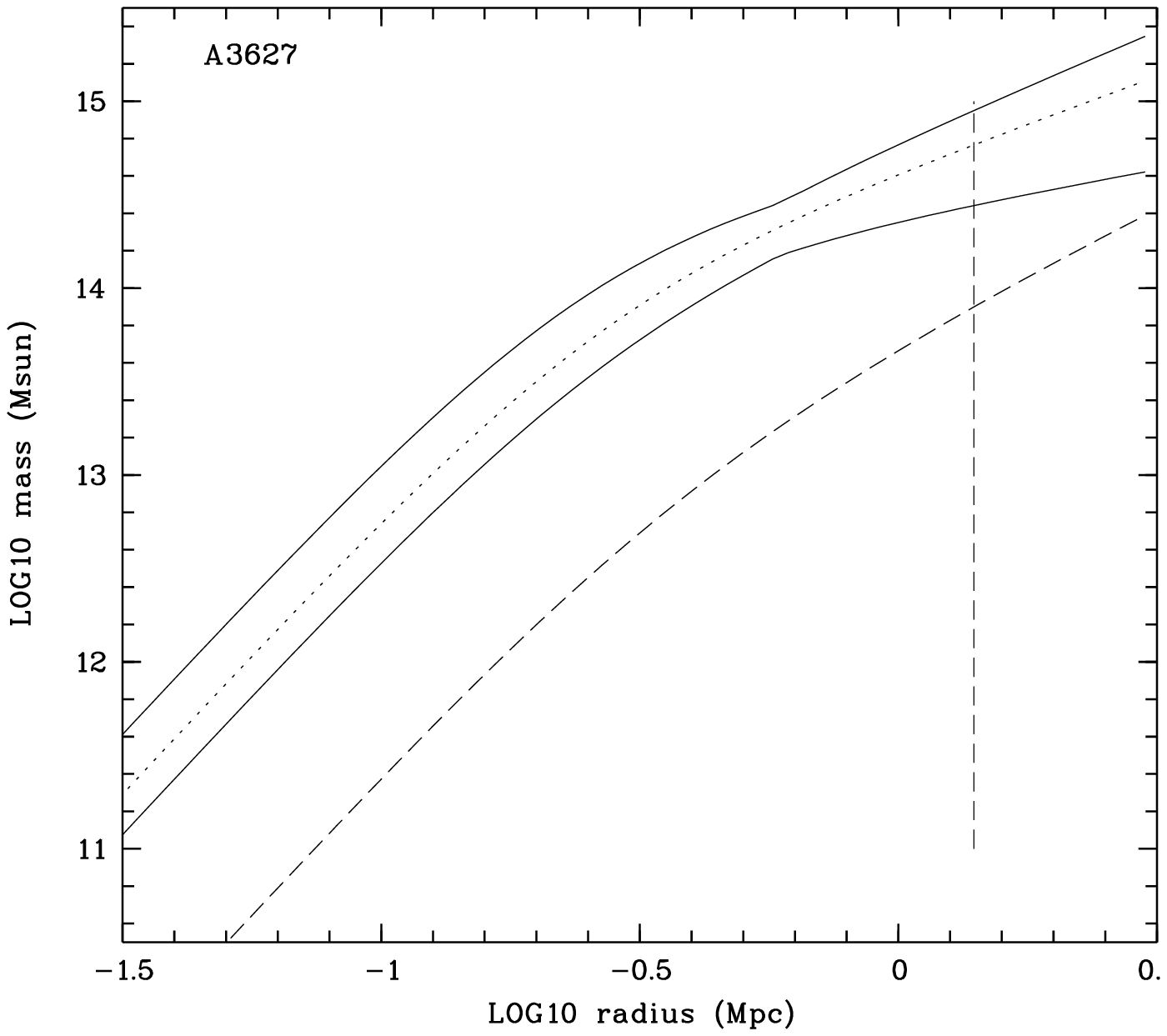,height=8cm}
\caption{Gravitational mass and gas mass profiles of A3627 determined from
the X-ray data. The solid curves show the upper and lower limit for the
gravitational mass profile calculated for intracluster medium temperatures
in the range 5.0 to 9.5 keV. The dotted curve shows the mass profile for an
isothermal model with a temperature of 7 keV. The dashed curve gives the gas
mass profile determined from the $\beta $ model fit to the surface
brightness profile. The dashed vertical line indicates the outer radius
to which X-ray emission from the cluster could be detected. The mass profiles
to the right of this line are only extrapolations.}
\end{figure}

\section {Discussion}

In the ROSAT All Sky Survey
A3627 is the sixth brightest cluster in the ROSAT band. The brightest
clusters in the survey are Virgo, Perseus, Coma, Ophiuchus, and Centaurus
with ROSAT PSPC count rates in the 0.5 to 2.0 energy band (channels 52
to 201) of $\ge$80, 38.3 ($\pm 0.3$), 16. ($\pm 0.2$), 11.5 ($\pm 0.3$),
and 9.3 ($\pm 0.2$) cts s$^{-1}$, respectively. A3627 follows
closely with a  ROSAT PSPC count rate of $7.0 (\pm  0.3)$ cts s$^{-1}$.
While the brightest 5 clusters are all very well known and have received
a lot of observational attention, e.g. with all recent X-ray
observatories, A3627 remained widely unnoticed.

Why has it been overlooked in previous X-ray surveys ?
For the calculated flux in the ROSAT energy band
of $2~ (\pm 0.2) \cdot 10^{-10}$ erg s$^{-1}$ cm$^{-2}$, we find an X-ray
flux in the 2 - 10 keV energy band of $1.6 - 2.8 \cdot 10^{-10}$
erg s$^{-1}$ cm$^{-2}$. The sensitivity limits of the
HEAO 1 survey are, for example, $1.9 \cdot 10^{-11}$ erg s$^{-1}$ cm$^{-2}$ for
the A2 instrument (e.g. Jahoda \& Mushotzky 1989) and $5 \cdot 10^{-12}$
erg s$^{-1}$ cm$^{-2}$ for the A1 instrument (e.g. Wood \et\ 1984).
Fig.8 shows a map of the HEAO 1 - A2
survey with a scale of 20 degrees around the cluster A3627 as obtained
from {\it SKYVIEW} (available on {\it World Wide Web}).
There is indeed a bright X-ray
source in the center of the image, labeled as no. 2
(no. 21 in Jahoda \& Mushotzky 1989), which is close to
the position of the cluster. This source, cataloged as 4U1556-60 or
1H1556-605, is identified with a galactic, low mass X-ray binary
(e.g. Motch \et\ 1989). The source count rate of 1H1556-605 in the HEAO 1 - A1
survey is $0.0446 (\pm 0.022)$ cts s$^{-1}$ corresponding to a 2-10 keV flux
of about $2 \cdot 10^{-10}$ erg s$^{-1}$ cm$^{-2}$ (Wood \et\ 1984).

Fig. 9 shows the region around A3627 in the ROSAT All Sky Survey with a scale
of about 4 by 4 degrees. 1H1556-605 is the bright source to the west at
a position of 16h~01~02, -60d44 17 (J2000) with a ROSAT PSPC count rate
in the hard band (channel 52 - 201) of about 8 cts s$^{-1}$. Thus the
flux of the low mass X-ray binary is only slightly higher in the ROSAT band
than the integrated X-ray flux from A3627. The fact that the
2 - 10 keV flux expected from A3627 is close to what is observed and that
the two sources appear about equally bright in the hard ROSAT band leads
to the conclusion that the flux measured for 1H1556-605 in the early
X-ray collimator observations contains a significant contribution from
the cluster emission. The distance between the two sources
is about 1.5 degrees. It is therefore
somewhat surprising that the X-ray source visible in the
HEAO 1 - A2 survey (cf. Fig. 8) is not more distorted compared to the appearance
of the other true point sources.

\begin{figure}[h]
\psfig{figure=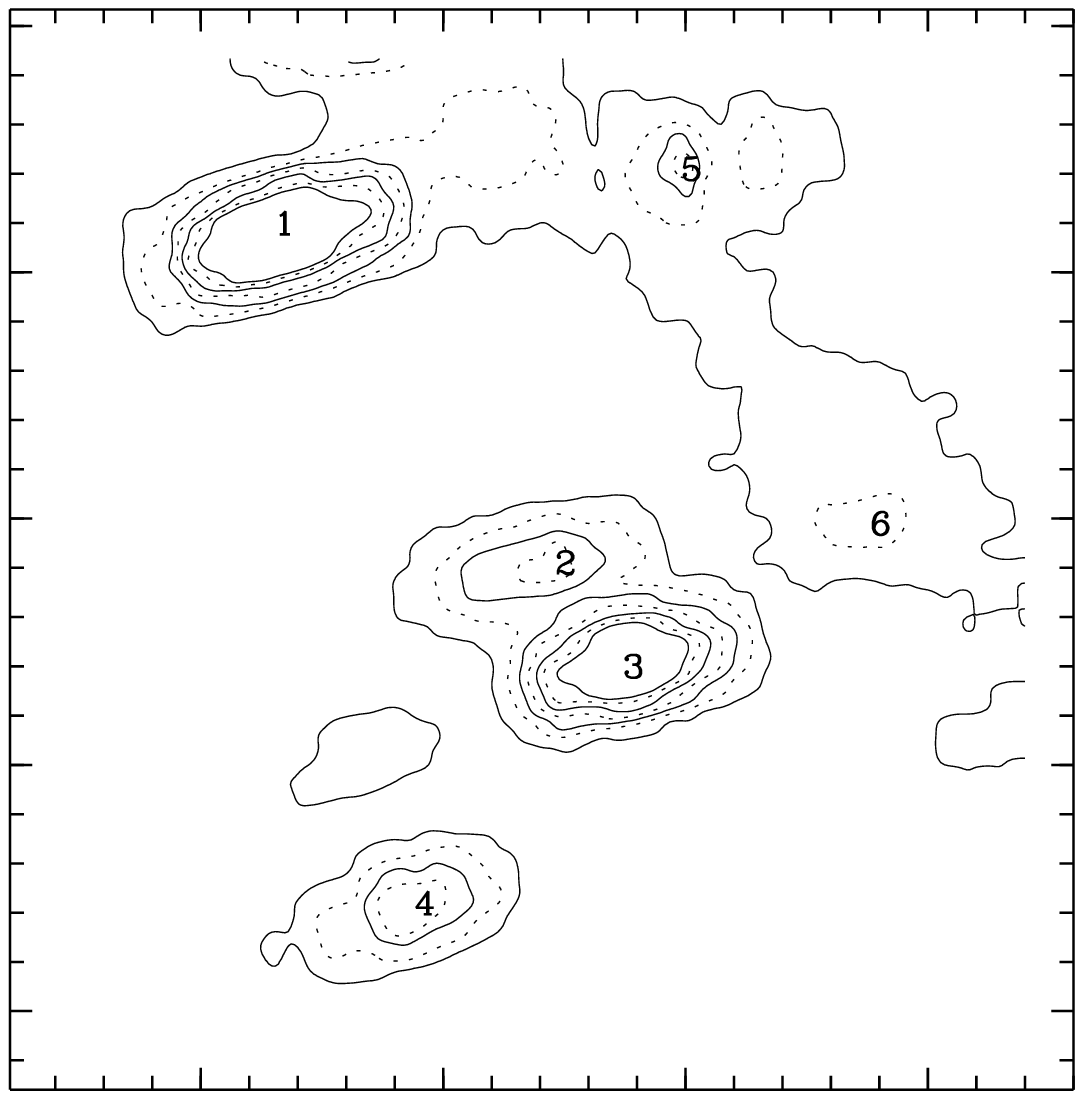,height=8cm}
\caption{X-ray emission from an area of 20 degree diameter around A3627 as
observed in the HEAO 1 - A2 All Sky Survey. A3627 is part of the source labeled
no. 2. The other sources are of galactic origin and are identified as:
1 = 4U1636-53, 3 = 1H1543-624, 4 = 1H1627-673, 5 = GPS1538-55, and 6 =
GPS 1510-58 (see also Jahoda \& Mushotzky 1989).}
\end{figure}

\begin{figure}[h]
\psfig{figure=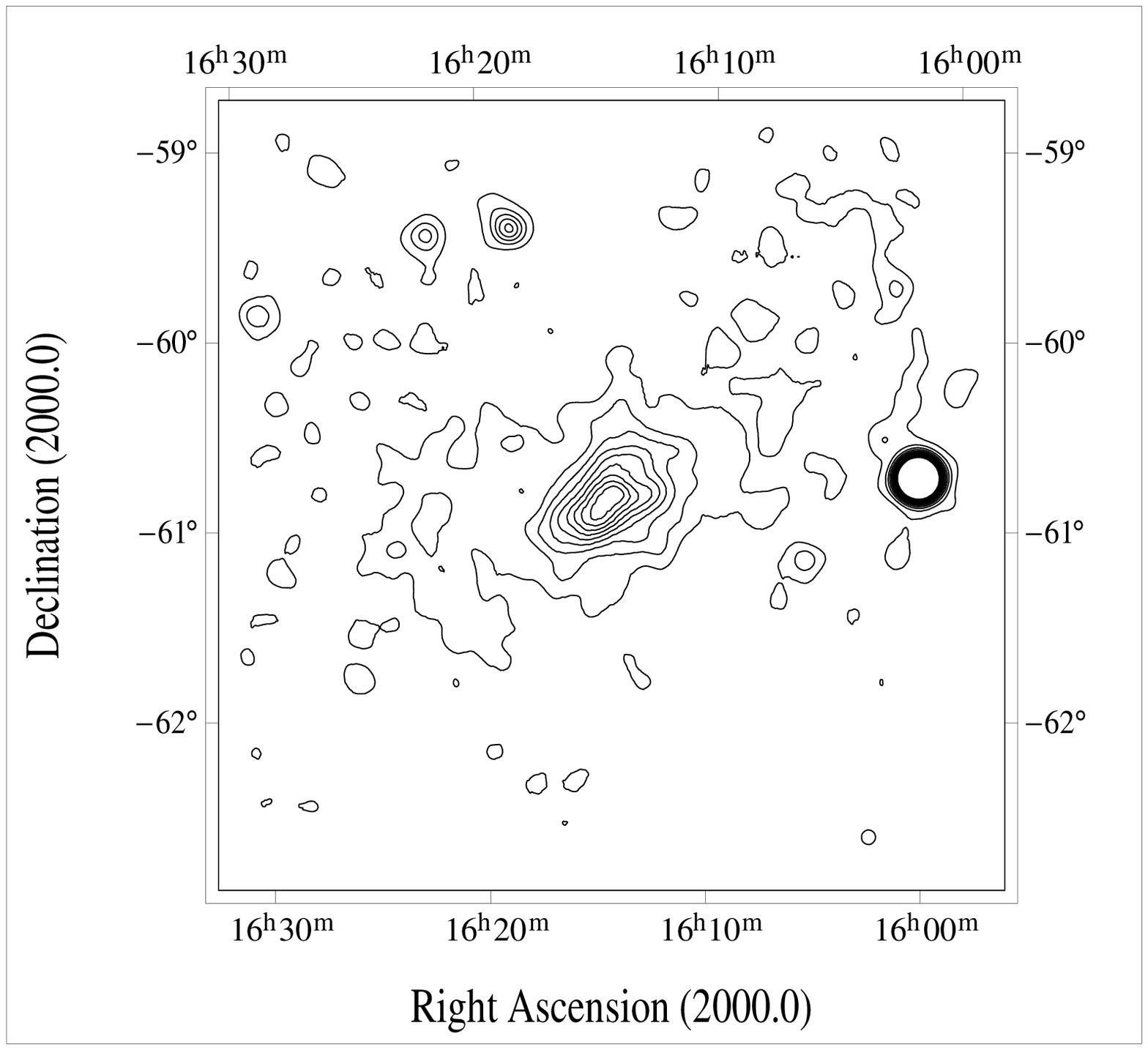,height=8cm}
\caption{X-ray map of the region around A3627 from the ROSAT All Sky Survey.
The image is produced from the photons in the 0.5 to 2 keV band.
The source in the center is
A3627 and the bright point source to the right is the low mass X-ray binary
1H1556-605. Both sources contribute to the HEAO 1 source labeled no. 2 in
Fig. 8. }
\end{figure}

The mass estimate above shows that A3627 has at least half the mass of
the Perseus and Coma clusters with masses of
$1.0 - 2.6 \cdot 10^{15}~h_{50}^{-1}$ \msu\ for an outer radius of
$3 ~h_{50}^{-1}$ Mpc (Briel \et\ 1991a, B\"ohringer 1994).
It is therefore a very prominent mass concentration in the local Universe.
Since it is located in the Great Attractor region it is responsible for
part of the gravitational pull causing the local streaming velocity
in this direction, as already discussed by Kraan-Korteweg \et\ (1996).

The mass obtained from the X-ray study can also be compared
to the mass estimate from a virial analysis based on
the optical data. For a velocity dispersion of 897 km s$^{-1}$
(corrected for measurement errors), a core radius
of $262 h_{50}^{-1}$ kpc, and an outer radius of $3 h_{50}^{-1}$ Mpc
we obtain a mass of $9.5 \cdot 10^{14} h_{50}^{-1}$~\msu using the formula (e.g.
Sarazin 1986):
\begin{equation}
M(<r) =  {9 \sigma_r^2~ r_c \over G} \left( ln(x+(1+x^2)^{1/2}) -
x(1+x^2)^{-1/2} \right)     ,
\end{equation}
where $x = r/r_c$. This result is in very
good agreement with the X-ray analysis. It implies that the X-ray
temperature estimate is consistent with the optical data. 

The mass of A3627 quoted in the
paper by Kraan-Korteweg \et\ (1996) with a value of
$5 \cdot 10^{15}$ \msu\ refers to a much larger volume. This mass was obtained
by integrating over a galaxy distribution described by a De Vaucouleur
profile (e.g. Rood \et\ 1972) with  effective radius
of 155 arcmin ($\sim 4.2 h_{50}^{-1}$ Mpc) with the same galaxy velocity
dispersion as used above. This mass therefore corresponds to a larger
structure and cannot easily be compared to our mass determination.
In addition the mass determined here
refers only to the spherically symmetric main part of the cluster.

A3627 obviously shows an interesting merger configuration in which a smaller
cluster component is merging with the major part in the south-eastern region.
The lower temperature found in the south-east
indicates that the smaller cluster component had an originally
lower gas temperature than the main cluster. This is consistent with the
fact that the cluster mass and the gas temperature are generally tightly
correlated (e.g. Edge \et 1991, David \et 1993)
and that one expects the second component of the cluster to
be significantly less massive than the main part from the morphological
appearance of the cluster. It also implies that the outer parts of the
infalling component have not been heated by the shock waves
which are expected from
the merger event (e.g. Schindler \& M\"uller 1993).

Two radio galaxies with extended radio lobes found in A3627 may give
further insight into the state of the merger. PKS1610-60.8 which
appears as a point source in the X-ray image is a wide-angle-tail (WAT) radio
galaxy (Christiansen \et\ 1977). The radio lobes show an extent of at least
8 arcmin (210 kpc) on both sides with bending angles of about $45 \deg$
at distances of 1.5 and 4 arcmin from the galactic center. This galaxy,
which is located approximately in the middle between the X-ray maxima
of the main component and the subcluster, is the 
the brightest galaxy of the cluster and of elliptical type
(see Christiansen \et 1977). The fact that this galaxy
is a WAT supports the assumption that it was the central galaxy
of the pre-merger main
cluster with a small relative velocity with respect to the overall
gravitational potential of the cluster. It is somewhat surprising, however,
that we don't see a stronger distortion of the radio lobes today, since the
galaxy seems to be located in the active region of the present merger.

The other interesting radio galaxy,
B1610-60.5 (Jones \& McAdam 1994), is located about 14 arcmin north-east
of the X-ray maximum. It is a head-tail radio source and has one of the
longest tails ever found with a size of $710 h_{50}^{-1}$ kpc (26 arcmin).
The tail extends in north-western direction (PA = $108 \deg$) and stays
within an angle of $10 \deg$ within this direction over the whole length.
It is striking that this direction almost coincides with the direction
of the line connection the X-ray maxima of the main part and subcomponent
of the cluster.
The straightness of the tail implies that the intracluster medium
was not disturbed by large flows during the passage of the galaxy (for
a time of about $7 \cdot 10^8$ years, assuming a galaxy velocity of about
1000 \kms). 

The lack of distortion of the radio lobes of the two radio galaxies and the
compactness of the subclump as it appears in Fig. 3 imply that the merger has
not progressed very far and that most of the main component of the cluster
is still undisturbed by the effects of the collision. Compared to A2256,
another famous merging cluster (Briel \et\ 1991b), the merging process
is closer to completion in A3627. A study of the
gas temperature distribution by future X-ray observations and the
combination of the present results with the redshift data from the galactic
plane galaxy survey will provide a better basis to elucidate the structure
of the merging clusters.

\section{Summary}

The low galactic latitude cluster A3627 ($b = 7\deg $) which received little
attention in the past is the 6$^{th}$ brightest cluster in the sky
in the ROSAT energy band. The X-ray flux is high enough that it has even
been seen in the Uhuru sky survey but it was confused with a galactic
low mass X-ray binary which lies close to it and has a similar X-ray flux.

A3627 has a mass around $10^{15}$ \msu. Therefore
it is almost as massive as the prominent Perseus and Coma clusters but, with
a redshift of $z =  0.016$, it is even closer.
The cluster is an interesting merger system and should be one of
the most interesting targets to study cluster mergers with spatially resolved
X-ray spectroscopy.

\acknowledgments

We thank the ROSAT team for providing the ROSAT All Sky Survey data for
the area around A3627 and for the excellent performance of the pointed
observation on the cluster. We also thank Patrick Woudt for the identification
of the X-ray point sources with individual galaxies. H.B. and S.S. are grateful
for financial support by the German BMFT through the Verbundforschung for
astronomy. S.S. gratefully acknowledges the hospitality of the Astronomical
Institute of the University of Basel.

%

\begin{planotable}{lrrr}
\tablewidth{27pc}
\tablecaption{Gravitational Mass and Gas Mass of A3627 }
\tablehead{
\colhead{radius $^{a)}$ } &              
\colhead{grav. mass $^{b)}$} &
\colhead{gas mass $^{c)}$} & 
\colhead{gas mass fraction (\%)}}

\startdata
1.0  & 2.2 - 5.8  & 0.46 & 8 - 21 $h_{50}^{-1.5}$ \nl
1.4  & 2.8 - 9.0  & 0.8  & 9 - 29 $h_{50}^{-1.5}$\nl
3.0  & 4.2 - 22.  & 2.45  &11 - 58 $h_{50}^{-1.5}$\nl
\tablenotetext{a}{~~radius in $h_{50}^{-1}$ Mpc}
\tablenotetext{b}{~~mass in $h_{50}^{-1} 10^{14}$ \msu}
\tablenotetext{c}{~~mass in $h_{50}^{-2.5} 10^{14}$ \msu}
\end{planotable}

\end{document}